\documentclass[10pt]{revtex4}
\newcount\driver
\newcount\bozza

\font\ottorm=cmr8\font\ottoi=cmmi8\font\ottosy=cmsy8%
\font\ottocss=cmcsc8%
\font\sixrm=cmr6\font\sixi=cmmi6\font\sixsy=cmsy6%
\font\fiverm=cmr5\font\fivesy=cmsy5
\font\fivei=cmmi5
\font\tenmib=cmmib10
\font\sevenmib=cmmib10 scaled 800

 2

\font\sc=cmcsc10

\font\elevenrm=cmr11
\font\twelverm=cmr12
\font\ottorm=cmr8
\font\msytw=msbm9 scaled\magstep1

\font\msytwww=msbm5 scaled\magstep1
\font\indbf=cmbx10 scaled\magstep2

\font\ottorm=cmr8\font\ottoi=cmmi8\font\ottosy=cmsy8%
\font\ottocss=cmcsc8%
\font\sixrm=cmr6\font\sixi=cmmi6\font\sixsy=cmsy6%
\font\fiverm=cmr5\font\fivesy=cmsy5
\font\fivei=cmmi5

\def\ottopunti{\def\rm{\fam0\ottorm}%
\textfont0=\ottorm\scriptfont0=\sixrm\scriptscriptfont0=\fiverm%
\textfont1=\ottoi\scriptfont1=\sixi\scriptscriptfont1=\fivei%
\textfont2=\ottosy\scriptfont2=\sixsy\scriptscriptfont2=\fivesy%
\textfont4=\ottocss\scriptfont4=\sc\scriptscriptfont4=\sc%
\scriptfont4=\ottocss\scriptscriptfont4=\ottocss%
\textfont5=\tenmib\scriptfont5=\sevenmib\scriptscriptfont5=\fivei
\setbox\strutbox=\hbox{\vrule height7pt depth2pt width0pt}%
\normalbaselineskip=9pt\let\sc=\sixrm\normalbaselines\rm}

\mathchardef\BDpr = "0540  
\mathchardef\Bg   = "050D  

{\count255=\time\divide\count255 by 60 \xdef\hourmin{\number\count255}
        \multiply\count255 by-60\advance\count255 by\time
   \xdef\hourmin{\hourmin:\ifnum\count255<10 0\fi\the\count255}}

\def\openone{\leavevmode\hbox{\elevenrm 1\kern-3.63pt\twelverm1}}%
\def\*{\vglue0.5truecm}

\let\a=\alpha \let\b=\beta  \let\g=\gamma  \let\d=\delta \let\e=\varepsilon
     \let\th=\theta  \let\l=\lambda
\let\m=\mu             \let\p=\pi    \let\r=\rho
\let\s=\sigma     \let\ph=\varphi
   
   \let\L=\Lambda

\def\\{\hfill\break} \let\==\equiv

\let\io=\infty 

\let\0=\noindent

\def\media#1{{\langle#1\rangle}}

\let\dpr=\partial

\def\tende#1{\,\vtop{\ialign{##\crcr\rightarrowfill\crcr
 \noalign{\kern-1pt\nointerlineskip} \hskip3.pt${\scriptstyle
 #1}$\hskip3.pt\crcr}}\,}
\def\circage{\lower2pt\hbox{$\,\buildrel > \over {\scriptstyle \sim}\,$}}
\def\otto{\,{\kern-1.truept\leftarrow\kern-5.truept\to\kern-1.truept}\,}

 \def\VV{{\cal V}}
 \def\HHH{{\cal H}}

\def\DD{{\cal D}} \def\SS{{\cal S}}

\def\T#1{{#1_{\kern-3pt\lower7pt\hbox{$\widetilde{}$}}\kern3pt}}
\def\VVV#1{{\VV #1}_{\kern-3pt
\lower7pt\hbox{$\widetilde{}$}}\kern3pt\,}
\def\W#1{#1_{\kern-3pt\lower7.5pt\hbox{$\widetilde{}$}}\kern2pt\,}

\def\indica{\leaders \hbox to 0.5cm{\hss.\hss}\hfill}
\def\guida{\leaders\hbox to 1em{\hss.\hss}\hfill}

   \def\qq{{\bf q}}
   \def\pp{{\bf p}}
 \def\xx{{\bf x}} \def\yy{{\bf y}} 
\def\hhh{{\bf h}}
\def\kk{{\bf k}}
\def\Vm{{\bf m}}\def\Vn{{\bf n}}

\def\VV#1{{\,\underline#1\,}}
\def\ul{\underline}

\mathchardef\aa   = "050B
\mathchardef\bb   = "050C
\mathchardef\ggg  = "050D
\mathchardef\xxx  = "0518
\mathchardef\zzzzz= "0510
\mathchardef\oo   = "0521
\mathchardef\lll  = "0515
\mathchardef\mm   = "0516
\mathchardef\Dp   = "0540
\mathchardef\H    = "0548
\mathchardef\FFF  = "0546
\mathchardef\ppp  = "0570
\mathchardef\Bn   = "0517
\mathchardef\pps  = "0520
\mathchardef\fff  = "0527
\mathchardef\FFF  = "0508
\mathchardef\nnnnn= "056E

\def\to{\rightarrow}

\def\qed{\hfill\raise1pt\hbox{\vrule height5pt width5pt depth0pt}}

\def\indic{\hbox{\raise-2pt \hbox{\indbf 1}}}

\def\RRR{\hbox{\msytw R}}

 \def\ZZZ{\hbox{\msytw Z}}
 \def\zzz{\hbox{\msytwww Z}}

\def\ul#1{{\underline#1}}

\def\V0{{\bf 0}}

\font\tenmib=cmmib10 
\font\sevenmib=cmmib7\font\fivemib=cmmib5 

\font\fivei=cmmi5\font\sixi=cmmi6\font\ottoi=cmmi8
\font\ottorm=cmr8\font\fiverm=cmr5\font\sixrm=cmr6
\font\ottosy=cmsy8\font\sixsy=cmsy6\font\fivesy=cmsy5
\font\ottocss=cmcsc8%

\mathchardef\Ba   = "050B  
\mathchardef\Bb   = "050C  
\mathchardef\Bg   = "050D  
\mathchardef\Bd   = "050E  
\mathchardef\Be   = "0522  
\mathchardef\Bee  = "050F  
\mathchardef\Bz   = "0510  
\mathchardef\Bh   = "0511  
\mathchardef\Bthh = "0512  
\mathchardef\Bth  = "0523  
\mathchardef\Bi   = "0513  
\mathchardef\Bk   = "0514  
\mathchardef\Bl   = "0515  
\mathchardef\Bm   = "0516  
\mathchardef\Bn   = "0517  
\mathchardef\Bx   = "0518  
\mathchardef\Bom  = "0530  
\mathchardef\Bp   = "0519  
\mathchardef\Br   = "0525  
\mathchardef\Bro  = "051A  
\mathchardef\Bs   = "051B  
\mathchardef\Bsi  = "0526  
\mathchardef\Bt   = "051C  
\mathchardef\Bu   = "051D  
\mathchardef\Bf   = "0527  
\mathchardef\Bff  = "051E  
\mathchardef\Bch  = "051F  
\mathchardef\Bps  = "0520  
\mathchardef\Bo   = "0521  
\mathchardef\Bome = "0524  
\mathchardef\BG   = "0500  
\mathchardef\BD   = "0501  
\mathchardef\BTh  = "0502  
\mathchardef\BL   = "0503  
\mathchardef\BX   = "0504  
\mathchardef\BP   = "0505  
\mathchardef\BS   = "0506  
\mathchardef\BU   = "0507  
\mathchardef\BF   = "0508  
\mathchardef\BPs  = "0509  
\mathchardef\BO   = "050A  
\mathchardef\BDpr = "0540  
\mathchardef\Bstl = "053F  

\def\V#1{{\bf#1}}
\let\aa=\Ba\let\fff=\Bf\def\HHH{{\cal H}}
\let\oo=\Bo\let\nn=\Bn
\let\pps=\Bps\def\hhh={\V h}
\let\bb=\Bb

\def\RRR{\hbox{\msytw R}}

 \def\ZZZ{\hbox{\msytw Z}}
 \def\zzz{\hbox{\msytwww Z}}

\let\ul=\underline

\def\ins#1#2#3{\vbox to0pt{\kern-#2 \hbox{\kern#1 #3}\vss}\nointerlineskip}
\newdimen\xshift \newdimen\xwidth \newdimen\yshift
\newcount\griglia

\def\insertplot#1#2#3#4#5#6{%
\begin{figure}[h]
\begin{center}
\vspace{#2pt}
\begin{minipage}{#1pt}
#3
\ifnum\driver=1
\griglia=#6
\ifnum\griglia=1
\openout13=griglia.ps
\write13{gsave .2 setlinewidth}
\write13{0 10 #1 {dup 0 moveto #2 lineto } for}
\write13{0 10 #2 {dup 0 exch moveto #1 exch lineto } for}
\write13{stroke}
\write13{.5 setlinewidth}
\write13{0 50 #1 {dup 0 moveto #2 lineto } for}
\write13{0 50 #2 {dup 0 exch moveto #1 exch lineto } for}
\write13{stroke grestore}
\closeout13
\includegraphics{griglia.ps}\fi
\includegraphics{#4.ps}\fi
\ifnum\driver=2
\fi
\end{minipage}
\end{center}
\caption{#5}
\end{figure}
}

\newdimen\shift \shift=-1truecm
\def\lb#1{%
\ifnum\bozza=1
\label{#1}\rlap{\kern\shift{$\scriptstyle#1$}}
\else\label{#1}
\fi}

\def\be{\begin{equation}}
\def\ee{\end{equation}}
\def\bea{\begin{eqnarray}}\def\eea{\end{eqnarray}}
\def\bean{\begin{eqnarray*}}\def\eean{\end{eqnarray*}}
\def\bfr{\begin{flushright}}\def\efr{\end{flushright}}
\def\bc{\begin{center}}\def\ec{\end{center}}
\def\ba#1{\begin{array}{#1}} \def\ea{\end{array}}
\def\bd{\begin{description}}\def\ed{\end{description}}
\def\bv{\begin{verbatim}}\def\ev{\end{verbatim}}
\def\nn{\nonumber}
\def\Halmos{\hfill\vrule height10pt width4pt depth2pt \par\hbox to \hsize{}}

\renewcommand{\theequation}{\arabic{section}.\arabic{equation}}

\newdimen\xshift \newdimen\xwidth \newdimen\yshift \newdimen\ywidth

\def\ins#1#2#3{\vbox to0pt{\kern-#2\hbox{\kern#1 #3}\vss}\nointerlineskip}

\def\eqfig#1#2#3#4#5{
\par\xwidth=#1 \xshift=\hsize \advance\xshift
by-\xwidth \divide\xshift by 2
\yshift=#2 \divide\yshift by 2
\line{\hglue\xshift \vbox to #2{\vfil
#3 \includegraphics{#4.ps}
}\hfill\raise\yshift\hbox{#5}}}

\def\8{\write12}

\begin{document}

\title{Long range order for lattice dipoles}
\*

\author{Alessandro Giuliani}
\affiliation{Dipartimento di Matematica, 
Universit\`a di Roma Tre, L.go S. Leonardo Murialdo 1, 00146 Roma Italy}
\date{25 March 2008}
\begin{abstract} We consider a system of classical Heisenberg spins on a 
cubic lattice in dimensions three or more, interacting via the dipole-dipole 
interaction. We prove that at low enough temperature the system displays 
orientational long range order, as expected by spin wave theory. 
The proof is based on reflection positivity methods. In particular, we 
demonstrate a previously unproven conjecture on the dispersion relation of the 
spin waves, first proposed by Fr\"ohlich and Spencer, which allows one to apply
infrared bounds for estimating the long distance behavior of the spin-spin 
correlation functions.
\end{abstract}

\maketitle

\section{Introduction}

Recent advances in film growth techniques, in the control of 
spin-spin interactions and in the ability to characterize
magnetic materials have revived interest in the low temperature physics of 
magnetic systems. Both experimental and theoretical studies have revealed 
several unusual properties of magnetic films, such as
spontaneous formation of striped patterns, reorientation 
transitions (in temperature and in the sample thickness), increase of the
static magnetization with increasing temperature, just to mention a few
\cite{DMW00}.
It is believed that an essential role in determining the nature and 
morphology of the ordered state is played by the dipolar 
interaction. Unfortunately, its long-range nature and its anisotropic 
character make many standard theoretical methods 
and numerical algorithms inapplicable. 
It is therefore not surprising that, for instance, 
existence of long range order in 2D continuous spin systems interacting 
via a pure dipole-dipole interaction at low temperatures 
is still a matter of discussion, even at 
a heuristic level. In fact, 
the case of 2D lattice dipoles is a paradigmatic example of 
a system where the Mermin-Wagner theorem cannot be applied, 
spin-wave theory does not provide any resolutive answer and  
numerical simulations are difficult because of the slow relaxation dynamics
associated with the long-range nature of the interaction. New and 
more sophisticated methods, such as renormalized spin wave theory 
\cite{DMBW97,CRRT00} or 
block spin reflection positivity \cite{GLL2}, 
are required to deal with this class of systems. 

At a rigorous level, even widely accepted results, such as the existence
of long range orientational order in three dimensional lattice dipole systems
or the existence of the infinite volume Gibbs state for any given 
domain shape are yet to be fully proved. Many fundamental contributions to 
the rigorous theory of lattice dipole systems (and more generally of 
dipole gases on the lattice or in the continuum) date back to the 80's.
The use of several different techniques, such as reflection 
positivity, correlation inequalities, cluster expansion, renormalization group,
allowed people to rigorously prove, e.g., no-screening theorems at any 
activity and temperature \cite{P79,FS}, analyticity of the pressure
at small activities \cite{GK83,BGN86,BGN89,BY90}, 
existence of a scaling limit (Gaussian 
free field) for lattice dipoles at small activities \cite{NS97}
and the existence of the thermodynamic limit in the continuum in 
three or more dimensions \cite{FP}, 
see \cite{BM} for a comprehensive review of these results. 
In a seminal paper, Fr\"ohlich and Spencer \cite{FS}, 
among several other results, 
proved existence of long range order for a system of {\it discrete} lattice
dipoles in two or more dimensions. They also described a proof for the 
physically relevant case of continuous dipoles on the cubic lattice in three 
or more dimensions. 
However, in the case of infinite-range interaction, they could not give a 
complete proof, and their argument was based on an unproven conjecture on the
dispersion relation of the spin waves. In this paper we will give a 
complete proof of long range order 
for classical dipoles on a cubic lattice in three or more dimensions,
proving in particular the aforementioned Fr\"ohlich-Spencer conjecture. 
The proof is based on reflection positivity and it extends ideas
proposed in \cite{FS}.

\section{Main results}

Let $\L$ be a periodic box in $\ZZZ^d$, viewed as the restriction of a 
periodic box in $\RRR^d$ to $\ZZZ^d$. We assume that 
$\L$ is of side $2L$, with $L$ even, and
we write: $\L=[-L+1,\ldots,L]^d$.
We consider the following Hamiltonian:
\be H_\L=\sum_{i,j=1}^d\;\sum_{\xx,\yy\in\L}
S^i_\xx W_{ij}(\xx-\yy) S^j_\yy
\label{1.1}\ee
where $\vec S_\xx$ is a unit vector and
$\{S^i_\xx\}_{i=1,\ldots,d}$ are its components. 
Moreover, denoting the Yukawa potential by 
\be Y_{\e}(\xx)=\int \frac{d\kk}{(2\p)^d}\frac{e^{i\kk\xx}}{\kk^2+
\e^2}\;,\label{1.1a}\ee
the interaction matrix $W(\xx)$ is defined as:
\be W_{ij}(\xx)=\sum_{\Vn\in\zzz^d}(-\dpr_i\dpr_j)Y_{\e_\L}(\xx+2\Vn L)\;, \qquad
\xx\neq\V0\label{1.2}\ee
and $W_{ij}(\V0)=\sum_{\Vn\neq\V0}(-\dpr_i\dpr_j)
Y_{\e_\L}(2\Vn L)$. The parameter $\e_\L$ 
is an infrared regulator that is sent to zero in the thermodynamic
limit, i.e., $\lim_{|\L|\to\io}\e_\L=0$. 

For any fixed $\L$ and $\b>0$, let us denote by $\media{\cdot}_{\b,\L}$ the 
Gibbs expectation given by the probability measure $Z_{\b,\L}^{-1}
\prod_{\xx\in\L}d\m(\vec S_\xx)e^{-\b H_\L}$, with $d\m(\vec S)$ the uniform 
measure on the unit sphere and $Z_{\b,\L}$ the obvious 
normalization factor. For any $\xx\in\L$, given the unit vector $\vec S_\xx$, 
we define:
\be \s^i_\xx=(-1)^{\xx+x_i}S^i_\xx\label{1.2a}\ee
and denote by $\vec\s_\xx$ the unit vector with components $\s_\xx^i$. 
Our main result is the following.\\
\\
{\bf Theorem 1 (Existence of orientational Long Range Order).}
{\it If $d\ge 3$, there exists $\b_d>0$ such that, if $\b>\b_d$, in the 
thermodynamic limit,
\be \lim_{|\L|\to\io}\frac1{|\L|^2}\sum_{\xx,\yy\in\L}\media{\vec\s_\xx\cdot
\vec\s_\yy}_{\b,\L}\ge c_d(\b)>0\;,\label{1.2b}\ee
with $c_d(\b)$ a suitable positive function, vanishing at $\b=\b_d$.}
\\
\\
Using the methods of \cite{FP}, it can be proved that the state 
$\media{\cdot}_{\b,\L}$ admits a thermodynamic limit, which we will denote by 
$\media{\cdot}_{\b}$. The theorem above implies that, for $\b>\b_d$, 
the infinite volume Gibbs state 
$\media{\cdot}_\b$ is {\it not} an extremal Gibbs state. From the assumed 
symmetry of $\media{\cdot}_{\b,\L}$ under exchanges of the coordinate axes,
and by the general theory of decomposition into extremal states,
it follows that $\media{\cdot}_\b$ is a mixture of at least $2d$ extremal 
Gibbs states (pure phases), $\media{\cdot}_\b^{(\l)}$, which break rotational 
invariance and are characterized by 
\be \media{S^i_\xx}_\b^{(\l)}=(-1)^{\xx+x_i}v^i_\l\;,\label{1.2e}\ee
where $\{\vec v_\l\,:\, \l=1,\ldots,2d,\ldots\}$ are vectors obtained 
from some vector $\vec v_0\in\RRR^d$ by applying arbitrary rotations around the
origin which leave the unit cube centered at the origin invariant.

The rest of the paper will be devoted to the proof of Theorem 1. 
As mentioned in
the introduction, we will follow the same strategy proposed by Fr\"ohlich 
and Spencer in \cite{FS} and we will extend their reflection positivity
method to prove, in particular, their Conjecture 7.9 \cite{FS}. 
In order to present a self-contained proof, we shall reproduce below some
of the statements already proved in \cite{FS}, including a construction
of the ground states of (\ref{1.1}). 

\section{Proof of Theorem 1}

\subsection{Reflection Positivity.}
Let us recall the notion of reflection positivity, adapted to the 
present case. If $\vec S$ is a unit vector, define
\be \big(R_i\vec S\big)^j=(-1)^{1-\d_{i,j}}S^j\;.\label{1.3}\ee
Let $\p_i$ be a pair of planes perpendicular to the $i$--th direction, midway 
in between two lattice planes and bisecting $\L$ into two pieces $\L_+$
and $\L_-$ of equal size. Let $r_i$ denote reflection of sites with respect
to $\p_i$. Clearly $r_i\L_-=\L_+$. We define
\be (\th_i\vec S)_\xx=R_i\vec S_{r_i\xx}\label{1.4}\ee
Let $\SS_\pm=\{\vec S_\xx\}_{\xx\in\L_\pm}$. If $A$ is a function of $\SS_+$
we set 
\be (\th_i A)(\SS_-)=A(\{(\th_i\vec S)_\xx\}_{\xx\in\L_-})\label{1.5}\ee
We shall say that the expectation $\media{\cdot}_{\b,\L}$ is reflection 
positive (RP) iff, for an arbitrary function $A$ of $\SS_+$ 
\be \media{\overline{\th_iA(\SS_-)}A(\SS_+)}_{\b,\L}\ge 0\label{1.6}\ee
for $i=1,\ldots,d$. 
As discussed in \cite{FILS1} and in \cite{FS}, a sufficient condition for 
(\ref{1.6}) to hold is the following. Let $\vec \r: \RRR^d\to\RRR^d$
be an arbitrary $\RRR^d$ valued function of $\RRR^d$.
Assume that 
\be-
\sum_{l,m=1}^d \int_{x_i>0\atop y_i<0}\r^l(\xx)(\th_i\r)^m(\yy)
(-\dpr_l\dpr_m)Y_{\e_\L}(\xx-\yy)\ge 0\label{1.7}\ee
for all $\vec \r(\xx)$ and $i=1,\ldots,d$. Then $\media{\cdot}_{\b,\L}$ is RP. 

In our context the proof of (\ref{1.7}) proceeds as follows. For definiteness,
let us assume that $i=1$. Let $x_1>0$
and let us rewrite 
\be Y_{\e_\L}(\xx)=\int \frac{d\kk}{(2\p)^d} \frac{e^{i\kk\xx}}{\kk^2+\e_\L^2}=
\frac1{2(2\p)^{d-1}}\int\frac{d\kk_\perp}{\sqrt{\kk_\perp^2+\e_\L^2}}
\,e^{i\kk_\perp\cdot\xx_\perp}e^{-|x_1|\sqrt{\kk_\perp^2+\e_\L^2}}\label{1.8}\ee
where in the last expression $\kk_\perp=(k_2,\ldots,k_d)$ and similarly for 
$\xx_\perp$. If $x_1>y_1$, given any two functions $\r_1(\xx),\r_2(\xx)$:
\bea&& \sum_{l,m=1}^d\r_1^l(\xx)\r_2^m(\yy)\dpr_l\dpr_m Y_{\e_\L}(\xx-\yy)
=\frac1{2(2\p)^{d-1}}\int\frac{d\kk_\perp}{\sqrt{\kk_\perp^2+\e_\L^2}}
\,e^{i\kk_\perp\cdot(\xx_\perp-\yy_\perp)}
e^{-(x_1-y_1)\sqrt{\kk_\perp^2+\e_\L^2}}\cdot\nn\\
&&\cdot
\left(\r_1^1(\xx)\sqrt{\kk_\perp^2+\e_\L^2}-i\kk_\perp\Br^\perp_1(\xx)\right)
\left(\r_2^1(\yy)\sqrt{\kk_\perp^2+\e_\L^2}-i\kk_\perp\Br^\perp_2(\yy)\right)
\label{1.9}\eea
where $\Br^\perp_1=(\r_1^2,\ldots,\r_1^d)$ and $\Br^\perp_2=
(\r_2^2,\ldots,\r_2^d)$.

Using this expression we find that, if $i=1$, (\ref{1.7}) can be rewritten as
\bea&& \frac1{2(2\p)^{d-1}}\int_{x_1>0}d\xx\,\int_{y_1>0}d\yy\,
\int\frac{d\kk_\perp}{\sqrt{\kk_\perp^2+\e_\L^2}}
\,e^{i\kk_\perp\cdot(\xx_\perp-\yy_\perp)}
e^{-(x_1+y_1)\sqrt{\kk_\perp^2+\e_\L^2}}\cdot\nn\\
&&\cdot
\left(\r^1(\xx)\sqrt{\kk_\perp^2+\e_\L^2}-i\kk_\perp\Br^\perp(\xx)\right)
\left(\r^1(\yy)\sqrt{\kk_\perp^2+\e_\L^2}+i\kk_\perp\Br^\perp(\yy)\right)
\label{1.10}\eea
that is clearly non-negative. Let us remark that condition (\ref{1.7}) is 
equivalent to the statement that 
$H_\L=H_\L(\SS_-,\SS_+)$ can be rewritten in the following form \cite{FILS1}:
\be H_\L(\SS_-,\SS_+)=H_+(\SS_+)+\th_iH_+(\SS_-)-\int d\r(\qq)\, 
\overline{\th_i C_\qq(\SS_-)} C_\qq(\SS_+) \;,\label{1.10a}\ee
for a suitable positive measure $d\r$. In our case, 
\be H_+(\SS_+)=\sum_{i,j=1}^d\;\sum_{\xx,\yy\in\L_+}
S^i_\xx W_{ij}(\xx-\yy) S^j_\yy\;.\label{1.10b}\ee
Moreover, if $i=1$ and $\L_+=[1,\ldots,L]\times [-L+1,\ldots,L]^{d-1}$,
$\qq$ is a $(d-1)$-dimensional vector and, defining ${\bf S}^\perp_\xx=(S^2_\xx,
\ldots,S^d_\xx)$, 
\bea&& C_{\qq}(\SS_+)=
\sum_{\xx\in\L_-}\Big[S^1_\xx\sqrt{\qq^2+\e_\L^2}-i\qq\cdot
{\bf S}^\perp_\xx \Big]e^{i\qq\cdot\xx_\perp}e^{-x_1\sqrt{\qq^2+\e_\L^2}}\nn\\
&& d\r(\qq)=\frac{d\qq}{2(2\p)^{d-1}}\frac
{e^{-\sqrt{\qq^2+\e_\L^2}}}{\sqrt{\qq^2+\e_\L^2}}\;.
\label{1.10c}\eea
If $i>1$ and/or $\L_+$ is different from $[1,\ldots,L]\times 
[-L+1,\ldots,L]^{d-1}$,
analogous expressions for $C_\qq$ and $d\r(\qq)$ will be valid. 

\subsection{Ground states.} 
Let us now show how reflection positivity allows us to construct the ground 
states of (\ref{1.1}). The key remark is that the positive measure
$d\r(\qq)$ in (\ref{1.10a}) induces the definition of a scalar product between 
spin configurations in $\L_+$. In particular, combining the Cauchy-Schwarz 
inequality and the inequality of arithmetic and geometric means,
we find that 
\bea&& \int d\r(\qq)\, 
\overline{\th_i C_\qq(\SS_-)} C_\qq(\SS_+)\le\nn\\
&&\qquad\qquad\le \Big[\int d\r(\qq)\, 
\overline{\th_i C_\qq(\th_i \SS_+)} C_\qq(\SS_+)\Big]^{1/2}\cdot \Big[
\int d\r(\qq)\, 
\overline{\th_i C_\qq(\SS_-)} C_\qq(\th_i \SS_-)\Big]^{1/2}\nn\\
&&\qquad\qquad\le 
\int d\r(\qq)\, 
\overline{\th_i C_\qq(\th_i \SS_+)} C_\qq(\SS_+)+ 
\int d\r(\qq)\, 
\overline{\th_i C_\qq(\SS_-)} C_\qq(\th_i \SS_-)\;.\label{1.10d}\eea
If we insert this estimate in (\ref{1.10a}), we find that 
\be H_\L(\SS_-,\SS_+)\ge \frac12 H_\L(\th_i\SS_+,\SS_+)+\frac12 H_\L(\SS_-,\th_i
\SS_-)\;,\label{1.10e}\ee
that is, either $\{\th_i\SS_+,\SS_+\}$ or $\{\SS_-,\th_i\SS_-\}$ has lower 
energy than $\{\SS_-,\SS_+\}$. 
If we keep reflecting in different planes, using 
the chessboard estimate (see Theorem 4.1 in \cite{FILS1}),
we find that the energy $H_\L(\SS)$ of
a generic configuration of spins $\SS=\{\vec S_\xx\}_{\xx\in \L}$ 
is bounded below by $|\L|^{-1}\sum_{\xx_0\in\L}H_\L(\SS_{\xx_0})$, where 
the spin at $\xx$ in the configuration $\SS_{\xx_0}$ is given by
$(-1)^\xx(-1)^{x_i}S^i_{\xx_0}$, $i=1,\ldots,d$. We now show that 
$H_\L(\SS_{\xx_0})$
is independent of $\vec S_{\xx_0}$. Note that this is not apriori obvious, 
since the Hamiltonian is not invariant under global rotations. Let 
$\vec S_{\xx_0}=(\sin\th\cos\phi,\sin\th\sin\ph,\cos\th)$.
Then 
\bea H_\L(\SS_{\xx_0})=-
\sum_{\xx\in\L\atop\yy\in\zzz^d\setminus\xx}(-1)^{\xx-\yy}&&\left\{
\sum_{i=1}^d (-1)^{x_i-y_i}(S_{\xx_0}^i)^2\dpr_i^2Y_{\e_\L}(\xx-\yy)+\right.\\
&&\left.+\sum_{i\neq 
j=1}^d(-1)^{x_i}(-1)^{y_j}S^i_{\xx_0}S^j_{\xx_0}\dpr_i
\dpr_jY_{\e_\L}(\xx-\yy)\right\}\nn\label{1.11}\eea
In (\ref{1.10}) the summations over $\xx$ and $\xx-\yy$ factor, both
for the term in the first line and for the one in the second line. 
In particular the term in the second line is zero after summation. 
So we are left with
\be H_\L(\SS_{\xx_0})=e_0|\L|\;,\qquad e_0=
\sum_{\xx\neq\V0}(-1)^{\xx+x_1}(-\dpr_1^2) 
Y_{\e_\L}(\xx)\label{1.12}\ee
where we used that $\sum_i(S_{\xx_0}^i)^2=1$ and the rotation symmetry of 
$Y_{\e_\L}(\xx)$.

\subsection{Infrared bounds.}
In this section we will extend the ideas used above to construct the 
ground states, and we will derive lower bounds on the Fourier transform of 
$\media{\vec\s_\xx\cdot\vec \s_\yy}_{\b,\L}$ (also known as {\it infrared 
bounds} \cite{FSS}). This will conclude the proof of Theorem 1. 
The main ingredients that we will need
are {\it Gaussian domination} \cite{FSS} and a refinement of the estimates
on the dispersion relation of the spin waves, including a proof
of Conjecture 7.9 in \cite{FS}. 

As a first step we map the spin system in (\ref{1.1}) onto a ferromagnetic spin
system via the mapping $\vec S_\xx\otto\vec \s_\xx$ defined by (\ref{1.2a}).
In terms of the $\vec\s$'s, the Hamiltonian can be rewritten as
\be H_\L'=\sum_{i,j=1}^d\sum_{\xx,\yy\in\L}
\s^i_\xx W_{ij}'(\xx,\yy) \s^j_\yy
\label{1.14}\ee
with $W_{ij}'(\xx,\yy)=(-1)^{\xx-\yy}(-1)^{x_i+y_j}W_{ij}(\xx-\yy)$. Note that
$W'(\xx,\yy)$ is not translation invariant. The discussion above shows that 
the ground states of $H'_\L$ are spin configurations with spins all pointing in
the same direction in space. Moreover $H'_\L$ is reflection positive with 
respect to the ferromagnetic reflection $(\th'_i\vec\s)_\xx=\vec\s_{r_i\xx}$.
We rewrite $H'_\L$ in the form
\be H'_\L=-\frac12\sum_{i,j=1}^d\sum_{\xx,\yy\in\L}
(\s^i_\xx-\s^i_\yy) W_{ij}'(\xx,\yy) (\s^j_\xx-\s^j_\yy)
+\sum_{i,j=1}^d\sum_{\xx\in\L}\s^i_\xx\s^j_\xx\sum_{\yy}W_{ij}'(\xx,\yy)
\label{1.15}\ee
Note that $\sum_{\yy}W_{ij}'(\xx,\yy)=e_0\openone$, so that 
\be H'_\L=-\frac12\sum_{i,j=1}^3\sum_{\xx,\yy\in\L}
(\s^i_\xx-\s^i_\yy) W_{ij}'(\xx,\yy) (\s^j_\xx-\s^j_\yy)
+e_0|\L|\label{1.16}\ee
Let us now define:
\be K_{\b,\L}(\ul{\vec h})=\langle\exp\Big
\{\frac\b{2}\sum_{i,j=1}^d\sum_{\xx,\yy\in\L}
(\s^i_\xx-\s^i_\yy-h^i_\xx+h^i_\yy) W_{ij}'(\xx,\yy) (\s^j_\xx-\s^j_\yy
-h^j_\xx+h^j_\yy)\Big\}\rangle_{0,\L}\label{1.17}\ee
for $\vec h_\xx$, $\xx\in\L$, real vectors. 
The chessboard estimate (see Theorem 4.1 in \cite{FILS1}) shows 
that $K_{\b,\L}(\ul{\vec h})\le K_{\b,\L}(\ul{\vec 0})=K_{\b,\L}$ (Gaussian 
domination). This implies 
$\frac{d^2}{d\l^2}K_{\b,\L}(\l\ul{\vec h})\big|_{\l=0}\le 0$, that is
\be \b\langle\big|\sum_{i,j=1}^d
\sum_{\xx,\yy\in\L}(h^i_\xx-h^i_\yy)W'_{ij}(\xx,\yy)(\s^j_\xx
-\s^j_\yy)\big|^2\rangle_{\b,\L}'
\le -\sum_{i,j=1}^d\sum_{\xx,\yy\in\L}(h^i_\xx-h^i_\yy)^*W'_{ij}(\xx,
\yy)(h^j_\xx-h^j_\yy)\label{1.18}\ee
where $\media{\cdot}_{\b,\L}'$ is the average with statistical weight 
$Z_{\b,\L}^{-1}
e^{-\b H'_\L}$. Note that (\ref{1.18}) holds apriori for real $\vec h_\xx$, 
but it extends to complex vectors \cite{FILS1}. 
In terms of the original spins, (\ref{1.18}) reads
\be 2\b\langle\big|\sum_{i,j=1}^d
\sum_{\xx,\yy\in\L}h^i_\xx W_{ij}(\xx-\yy)S^j_\yy-e_0\sum_{\xx\in\L}\vec h_\xx
\cdot\vec S_\xx\big|^2\rangle_{\b,\L}
\le \sum_{i,j=1}^d\sum_{\xx,\yy\in\L}(h^i_\xx)^* W_{ij}(\xx-\yy)
h^j_\yy -e_0\sum_{\xx\in\L}|\vec h_\xx|^2\label{1.19}\ee
for some new vectors $\vec h_\xx$ (that we are denoting by the same symbol
for simplicity). Let $\pp \in\DD_L\=\{\frac{2\p}{2L}\Vm\,,\ 0\le m_1,m_2,m_3<2L
\}$ and 
let $\hat W_{ij}(\pp)=\sum_{\xx\in\L}e^{i\pp\xx}W_{ij}(\xx)$ denote the 
Fourier transform of $W(\xx)$. We shall also define 
\be \hat S^i_\pp=\frac1{|\L|^{1/2}}\sum_{\xx\in\L}e^{i\pp\xx} S^i_\xx\;,\qquad
Q_{ij}(\pp)=\media{\hat S^i_\pp\hat S^j_\pp}_{\b,\L}\;.\label{1.19a}\ee
Given $\pp$ such that $\hat W^0(\pp)\=
\hat W(\pp)-e_0\ge 0$ as 
a matrix and choosing $\vec h_\xx=|\L|^{-1/2}
e^{-i\pp\xx}\vec v$ in (\ref{1.19}), we get:
\be 0\le\hat W^0(\pp)\,Q(\pp)\,\hat W^0(\pp)\le \frac1{2\b}
\hat W^0(\pp)\;,\label{1.20aa}
\ee
in the sense of an inequality between non-negative matrices. If, moreover,
$\hat W^0(\pp)> 0$, then
\be 0\le Q(\pp)\le \frac1{2\b}\hat W^0(\pp)^{-1}\;.\label{1.20}\ee
Eq.(\ref{1.20aa})-(\ref{1.20}) are the key bounds. In order to make use of 
them, we need to study some properties of $\hat W^0(\pp)$, in particular we 
need to:
(i) show that $\hat W^0
\ge 0$; (ii) determine the set $S$ 
of momenta where $\hat W^0(\pp)$ has a vanishing eigenvalue; (iii)
determine the behavior of its eigenvalues close to the zeros. 
Property (i) was prover in \cite{FS} (we shall reproduce the proof below).
Moreover, in \cite{FS} it was conjectured (see Conjecture 7.9 in \cite{FS})
that, if $\Bp_\ell$ is the vector with components $\big(\Bp_\ell\big)^j=
\pi(1-\d_{\ell,j})$, then $S=\{\Bp^{(\ell)}\,:\,\ell=1,\ldots,d\}$ and the 
eigenvalue $\l_\ell(\pp)$ 
vanishing at $\Bp_\ell$, satisfies $\l_\ell(\pp)\ge c |\pp-\Bp_\ell|^2$
close to $\Bp_\ell$, for some $c>0$. 
Our next goal will be to prove this conjecture.

Let us start with showing that $\hat W^0(\pp)\ge 0$, for all $\pp$'s. 
This is equivalent to the claim that, for any $\vec v\in\RRR^d$ and for 
$\vec h_\xx=e^{i\pp\xx}\vec v$, 
\be \sum_{i,j=1}^d\sum_{\xx,\yy\in\L}(h^i_\xx)^* W^0_{ij}(\xx-\yy) h^j_\yy\ge 0
\;.\label{1.21}\ee
By the chessboard estimate, the left hand side of 
(\ref{1.21}) is bounded below by 
$|\L|^{-1}\sum_{\xx_0\in\L}(H_\L(\HHH_{\xx_0})-e_0|\vec v|^2)$, where 
the spin at $\xx$ in the configuration $\HHH_{\xx_0}$ is given by
$(-1)^\xx(-1)^{x_i}h^i_{\xx_0}$, $i=1,\ldots,d$. Now note that $H_\L(\HHH_{\xx_0})=
e_0$, so the proof of (\ref{1.21}) is concluded. Note that if $\pp\in S$ then 
$\hat W^0(\pp)=0$. We now want to get a more refined bound from below on 
$\hat W^0(\pp)$. In order to do this, we use the assumption that $L$ is even 
(so that the side of $\L$ is divisible by 4) and we repeatedly reflect the 
left hand side of (\ref{1.21}) in planes $\p_1$ bisecting the 
horizontal bonds of the form $\{(2m-1,\xx_\perp),(2m,\xx_\perp)\}$.
Finally we repeatedly reflect in all possible planes $\p_2,\ldots,\p_d$. 
The result is a bound from below of the form:
\be \sum_{i,j=1}^d v^i\hat W^0_{ij}(\pp)v^j\ge \frac1{|\L|}\sum_{i,j=1}^d 
\sum_{\xx,\yy\in\L}(u^i_\xx)^* W_{ij}(\xx-\yy) u^j_\yy-e_0|\vec v|^2
\;,\label{1.22}\ee
with
\bea&& u_\xx^1=e^{i\frac{p_1}2 f_1(x_1)}(-1)^{\xx+x_1}\;v_1\;,\nn\\
&&u_\xx^i=e^{i\frac{p_1}2 f_1(x_1)}
(-1)^{\xx+x_1+x_i}f_0(x_1)v_i\;,\qquad\quad i>1\;,\label{1.22a}\eea
where:
\be f_0(x)=\cases{1\ &if\ \ $x=4k$\cr 1\ &if\ \ $x=4k+1$\cr -1\ &if\ \ 
$x=4k+2$\cr -1\ &if\ \ $x=4k+3$\cr}\;,\qquad\quad f_1(x)=f_0(x-1)\label{1.23}\ee
and $k\in\ZZZ$. Given a lattice function $g(x)$ 
of a single variable with period 4, we shall write $g(x)=\{g(0),g(1),g(2),g(3)
\}$. Note that with this convention $f_0(x)=\{1,1,-1,-1\}$ and $f_1(x)=\{-1,1,
1,-1\}$. 

We now turn to the computation of the r.h.s. of (\ref{1.22}).
First of all, note that $\sum_{\xx,\yy\in\L}(u^i_\xx)^* W_{ij}(\xx-\yy) 
u^j_\yy=0$ if $i\neq j$. This is because the double summation
can be rewritten in the form $\sum_{\xx\in\L}\sum_{\yy\in\zzz^d\setminus\xx}
F_{ij}(\xx,\xx-\yy)$, for some function $F_{ij}(\xx,\xx-\yy)$ that is odd
in $(x_2-y_2)$ and/or in $(x_3-y_3), \ldots, (x_d-y_d)$, 
depending on the specific matrix element.
Let us consider $\sum_{\xx,\yy\in\L}(u^1_\xx)^* W_{11}(\xx-\yy) 
u^1_\yy$. This double summation is equal to
\be v_1^2 \sum_{\xx\in\L}\sum_{\yy\in\zzz^d\setminus\xx}
e^{-i\frac{p_1}{2}(f_1(x_1)-f_1(y_1))}(-1)^{\xx-\yy+(x_1-y_1)}
(-\dpr^2_1)Y_{\e_\L}(\xx-\yy)\label{1.24}\ee
A computation shows that $f_1(x)-f_1(y)=f_1(x)[1-g_0(x-y)]+f_0(x)g_1(x-y)$,
with $g_0(x)=\{1,0,-1,0\}$ and $g_1(x)=g_0(x-1)$. Then we can rewrite 
(\ref{1.24}) as
\be v_1^2 \sum_{\xx\in\L}\sum_{\yy\in\zzz^d\setminus\xx}
e^{-i\frac{p_1}{2}\big\{f_1(x_1)[1-g_0(x_1-y_1)]+f_0(x_1)g_1(x_1-y_1)\big\}}
(-1)^{\xx-\yy+(x_1-y_1)}
(-\dpr^2_1)Y_{\e_\L}(\xx-\yy)\label{1.25}\ee
Performing the summation over $\xx$ yields
\bea&& v_1^2 |\L|\sum_{\xx\neq\V0}\cos\left(\frac{p_1}2(1-g_0(x_1))\right)
\cos\left(\frac{p_1}2 g_1(x_1)\right)(-1)^{\xx+x_1}
(-\dpr^2_1)Y_{\e_\L}(\xx)=\nn\\
&& =v_1^2 |\L|\Big[e_0-\sin^2\frac{p_1}2\sum_{\xx\neq\V0}(1-g_0(x_1))
(-1)^{\xx+x_1}(-\dpr^2_1)Y_{\e_\L}(\xx)\Big]\label{1.26}\eea
The conclusion is 
\be \frac1{|\L|}\sum_{\xx,\yy\in\L}(u^1_\xx)^* W_{11}^0(\xx-\yy)u^1_\yy=
\a
\sin^2\frac{p_1}2\label{1.27}\ee
where $\a=\sum_{\xx\neq\V0}(1-g_0(x_1))
(-1)^{\xx+x_1}\dpr^2_1Y_{\e_\L}(\xx)$. 
We will show in Appendix A that $\a>0$.
Let us now consider $\sum_{\xx,\yy\in\L}(u^2_\xx)^* 
W_{22}(\xx-\yy) u^2_\yy$. This double summation is equal to
\be v_2^2 \sum_{\xx\in\L}\sum_{\yy\in\zzz^d\setminus\xx}
e^{-i\frac{p_1}{2}(f_1(x_1)-f_1(y_1))}f_0(x_1)f_0(y_1)(-1)^{(x_3-y_3)+\cdots+(x_d-y_d)}
(-\dpr^2_2)Y_{\e_\L}(\xx-\yy)\label{1.28}\ee
Using that $f_1(x)f_1(y)=g_0(x-y)+(-1)^xg_1(x-y)$ and $f_0(x)f_0(y)
=g_0(x-y)-(-1)^xg_1(x-y)$, we can rewrite this as:
\bea&& v_2^2 \sum_{\xx\in\L}\sum_{\yy\in\zzz^d\setminus\xx}
e^{-i\frac{p_1}{2}\big\{f_1(x_1)[1-g_0(x_1-y_1)]+f_0(x_1)g_1(x_1-y_1)\big\}}
\big[g_0(x_1-y_1)-(-1)^{x_1}g_1(x_1-y_1)\big]\cdot\nn\\
&&\hskip6.truecm\cdot(-1)^{(x_3-y_3)+\cdots+(x_d-y_d)}
(-\dpr^2_2)Y_{\e_\L}(\xx-\yy)\label{1.29}\eea
Performing summation over $\xx$ yields
\bea&& v_2^2 |\L|\sum_{\xx\neq\V0}\Bigg[g_0(x_1)
\cos\left(\frac{p_1}2(1-g_0(x_1))\right)
\cos\left(\frac{p_1}2 g_1(x_1)\right)-\\
&&\hskip1.5truecm-g_1(x_1)\sin\left(\frac{p_1}2
(1-g_0(x_1))\right)\sin\left(\frac{p_1}2 g_1(x_1)\right)\Bigg]
(-1)^{x_3+\cdots+x_d}(-\dpr^2_2)Y_{\e_\L}(\xx)\nn\label{1.30}\eea
This can be rewritten as
\be v_2^2 |\L|\Big[e_0-\cos^2\frac{p_1}2\sum_{\xx\neq\V0}(1-g_0(x_1))
(-1)^{\xx+x_2}(-\dpr^2_2)Y_{\e_\L}(\xx)\Big]\label{1.31}\ee
The conclusion is 
\be \frac1{|\L|}\sum_{\xx,\yy\in\L}(u^2_\xx)^* W_{22}^0(\xx-\yy)u^2_\yy=
\g\cos^2\frac{p_1}2\label{1.32}\ee
where $\g=\sum_{\xx\neq\V0}(1-g_0(x_1))
(-1)^{\xx+x_2}\dpr^2_2Y_{\e_\L}(\xx)$. We will show in Appendix A that $\g>0$.
Finally (\ref{1.32}) is valid even if in the l.h.s we exchange the index $2$ 
with $j\ge 3$. Substituting all this into (\ref{1.22}) yields:
\be \hat W^0(\pp)\ge \pmatrix{\a\sin^2\frac{p_1}2&0&0&0\cr
0&\g\cos^2\frac{p_1}2&0&0\cr
0&0&\ddots&0 \cr
0&0&0& \g\cos^2\frac{p_1}2}\;.\label{1.33a}\ee
%
By interchanging the roles
of $p_1,\ldots,p_d$ we finally get
\be \hat W^0(\pp)\ge Y(\pp)\;,\qquad
Y_{ij}(\pp)=\frac1d
\d_{ij}\Big(\a\sin^2\frac{p_i}2+\g\sum_{l\neq i}\cos^2\frac{p_l}2
\Big)\label{1.33}\ee
This bound proves Conjecture 7.9 in \cite{FS}. 

Now, the proof of existence of long range order, given the bounds (\ref{1.20}) 
and (\ref{1.33}), is standard. We reproduce it here, for completeness. 
Given $\ell\in\{1,\ldots,d\}$, we note that, by the rotation symmetry
of $\media{\cdot}_{\b,\L}$, 
\be \frac1{|\L|}\sum_\pp\media{\hat S^\ell_\pp\hat S^\ell_{-\pp}}_{\b,\L}=\frac1d
\;.\label{1.34}\ee
Therefore,
\be \frac1{|\L|}\media{\hat S^\ell_{\Bp_\ell}\hat S^\ell_{-\Bp_\ell}}_{\b,\L}=\frac1d
-\frac1{|\L|}\sum_{\pp\neq\Bp_\ell} Q_{\ell\ell}(\pp)\;.\label{1.35}\ee
Let us now consider the sum in the r.h.s.
Note that, for any $\pp\not\in S$, we can use (\ref{1.20}) and (\ref{1.33})
to conclude that 
\be Q_{\ell\ell}(\pp)\le \frac{d}{2\b}\Big(\a\sin^2\frac{p_\ell}2+\g
\sum_{j\neq \ell}\cos^2\frac{p_j}2\Big)^{-1}\;.\label{1.36}\ee
On the other hand, for $\pp=\Bp_m$, $m\neq\ell$, we can note that 
$\hat W^0_{ij}(\Bp_m)=\d_{ij}(1-\d_{im})(e_1-e_0)$, with $e_1=
\sum_{\xx\neq\V0}(-1)^{\xx+x_2}(-\dpr_1^2) 
Y_{\e_\L}(\xx)$ and $e_0$ defined in (\ref{1.12}). Using (\ref{1.33}), we
see that $e_1-e_0\ge(\a+\g)/d$. Therefore, if $v^i=\d_{i\ell}/(e_1-e_0)$,  
we can use (\ref{1.20aa}) to get
\be \sum_{i,j}v^i\Big[\hat W^0(\Bp_m)\,Q(\Bp_m)\,\hat W^0(\Bp_m)\Big]_{ij}v^j
= Q_{\ell\ell}(\Bp_m)\le \frac1{2\b}\frac1{e_1-e_0}\le 
\frac{d}{2\b}\frac1{\a+\g}
\label{1.37}\ee
and we conclude that (\ref{1.36}) is valid for $\pp=\Bp_m$, 
$m\neq\ell$, as well. Substituting (\ref{1.36}) we get
\be \frac1{|\L|}\media{\hat S^\ell_{\Bp_\ell}\hat S^\ell_{-\Bp_\ell}}_{\b,\L}\ge
\frac1d-\frac{d}{2\b|\L|}\sum_{\pp\neq\Bp_\ell}\frac1{\a\sin^2\frac{p_\ell}2+\g
\sum_{j\neq \ell}\cos^2\frac{p_j}2}\;.\label{1.38}\ee
If we note that 
\be \media{\hat S^\ell_{\Bp_\ell}\hat S^\ell_{-\Bp_\ell}}_{\b,\L}=\frac1{|\L|}
\sum_{\xx,\yy\in\L}\media{\s^\ell_\xx\s^\ell_\yy}_{\b,\L}\label{1.39}\ee
and take the thermodynamic limit in (\ref{1.38}) we finally get 
(\ref{1.2b}), with 
\be c_d(\b)=1-\frac{d^2}{2\b}\int_{[-\p,\p]^d}\frac{d\pp}{(2\p)^d}
\frac1{\a\sin^2\frac{p_\ell}2+\g
\sum_{j\neq \ell}\cos^2\frac{p_j}2}\;.\label{1.40}\ee

\acknowledgments I am grateful to Joel L. Lebowitz and Elliott H. Lieb
for stimulating my interest for this problem and for many helpful discussions.
Part of this work was supported by the Department of Physics of Princeton 
University and by U.S. National Science Foundation
grant PHY-0652854, which are gratefully acknowledged.

\appendix
\section{The constants $\a$ and $\g$ are positive
}\lb{A}
\setcounter{equation}{0}
\renewcommand{\theequation}{\ref{A}.\arabic{equation}}

In this Appendix we show that the constants $\a$ and $\g$ in (\ref{1.33}),
defined right after (\ref{1.27}) and (\ref{1.32}) respectively, are positive. 
Let us first consider $\a=\sum_{\xx\neq\V0}(1-g_0(x_1))
(-1)^{\xx+x_1}\dpr^2_1Y_{\e_\L}(\xx)$. Note that $1-g_0(x_1)$ is a 
non-negative even function of
$x_1$, vanishing at $x_1=0$. Using (\ref{1.8}) we can rewrite
\be \a=\frac1{(2\p)^{d-1}}
\sum_{x_1>0\atop \xx_\perp\in\zzz^{d-1}} (1-g_0(x_1))\int
d\kk_\perp\sqrt{\kk_\perp^2+\e_\L^2}
\,e^{i[\kk_\perp+\p{\bf 1}]\cdot\xx_\perp}e^{-x_1\sqrt{\kk_\perp^2+\e_\L^2}}\;,
\label{A.1}\ee
where $\p{\bf 1}$ is the $(d-1)$-dimensional vector
with components all equal to $\p$.
The summation over $\xx_\perp$ can be explicitly computed and produces
a $(d-1)$-dimensional delta function $(2\p)^{d-1}\sum_{\Vm\in\zzz^{d-1}}
\d(\kk_\perp+\p(2\Vm+{\bf 1}))$. So we find
\be \a=\sum_{x_1>0}\sum_{\Vm\in\zzz^d}
(1-g_0(x_1))\sqrt{\p^2(2\Vm+{\bf 1})^2+\e_\L^2}
\,e^{-x_1\sqrt{\p^2(2\Vm+{\bf 1})^2+\e_\L^2}}>0
\label{A.2}\ee
and the proof that $\a>0$ is concluded.

Similarly, let us consider $\g=\sum_{\xx\neq\V0}
(1-g_0(x_1))(-1)^{\xx+x_2}\dpr^2_2Y_{\e_\L}(\xx)$. Note that $(-1)^{x_1}(1-
g_0(x_1))$ is an even function of $x_1$, vanishing at $x_1=0$ with 
alternating signs. Using (\ref{1.8}) we can rewrite
\be \g=\frac1{(2\p)^{d-1}}
\sum_{x_1>0\atop\xx_\perp\in\zzz^{d-1}}(-1)^{x_1+1}(1-g_0(x_1))
\int\frac{d\kk_\perp}{\sqrt{\kk_\perp^2+\e_\L^2}}
\,k_2^2e^{ik_2x_2}e^{-x_1\sqrt{\kk_\perp^2+\e_\L^2}}\prod_{j\ge 3}e^{i(k_j+\p)x_j}
\label{A.3}\ee
Performing the summation over $\xx_\perp$ and using that 
$(-1)^{x_1+1}(1-g_0(x_1))=\{0,1,-2,1\}$, we find
\be\g=\sum_{\Vm\in\zzz^{d-1}}\frac{4\p^2m_1^2}{\sqrt{\p^2
(2\Vm+\p{\bf 1}-\hat e_1)^2+\e_\L^2}}\,\frac{\cosh\sqrt{\p^2
(2\Vm+\p{\bf 1}-\hat e_1)^2+\e_\L^2}-1}{\sinh2\sqrt{\p^2
(2\Vm+\p{\bf 1}-\hat e_1)^2+\e_\L^2}}
\label{A.4}\ee
where $\hat e_1=(1,0,\ldots,0)$. The proof that $\g>0$ is concluded.

\end{document}